\documentclass[prl,twocolumn,showpacs,10pt]{revtex4}

\usepackage{graphicx}
\usepackage{dcolumn}
\usepackage{bm}
\usepackage{times}

\begin{document}


\title{ 
A new polymorphic material?
Structural degeneracy of ZrMn$_2$ 
}

\author{Xing-Qiu Chen$^{a,c}$}
\author{W.Wolf$^b$}
\author{R.Podloucky$^a$}
\author{P.Rogl$^a$}
\author{M.Marsman$^d$}

\affiliation{%
$^a$ Institut f\"ur Physikalische Chemie, Universit\"at Wien,
 Liechtensteinstrasse 22A, A 1090, Vienna, Austria,}%
\affiliation{%
$^b$ Materials Design s.a.r.l., 44.av.F.-A. Bartholdi, 7200 Le Mans, France,
}%
\affiliation{%
$^c$ School of Materials and Metallurgy, Northeastern University, Shenyang
110004, Peoples' Republic of China, 
}
\affiliation{%
$^d$ Institut f\"ur Materialphysik, Universit\"at Wien, Sensengasse 8, A 1090, Vienna, Austria.}
\date{\today}
\begin{abstract}
Based on density functional calculations, we propose that ZrMn$_2$ is a
polymorphic material. We predict that at low temperatures the cubic C15, and
the hexagonal C14 and C36 structures of the Laves phase compound ZrMn$_2$ are
nearly equally stable within 0.3 kJmol$^{-1}$ or 30 K.  This degeneracy occurs
when the Mn atoms magnetize spontaneously in a ferromagnetic arrangement
forming the states of lowest energy.  From the temperature dependent free
energies at T $\approx$ 160K we predict a transition from the most stable C15
to the C14 structure, which is the experimentally observed structure at
elevated temperatures.
\end{abstract}
\maketitle
\begin{figure}
\begin{center}
 \includegraphics[width=85mm]{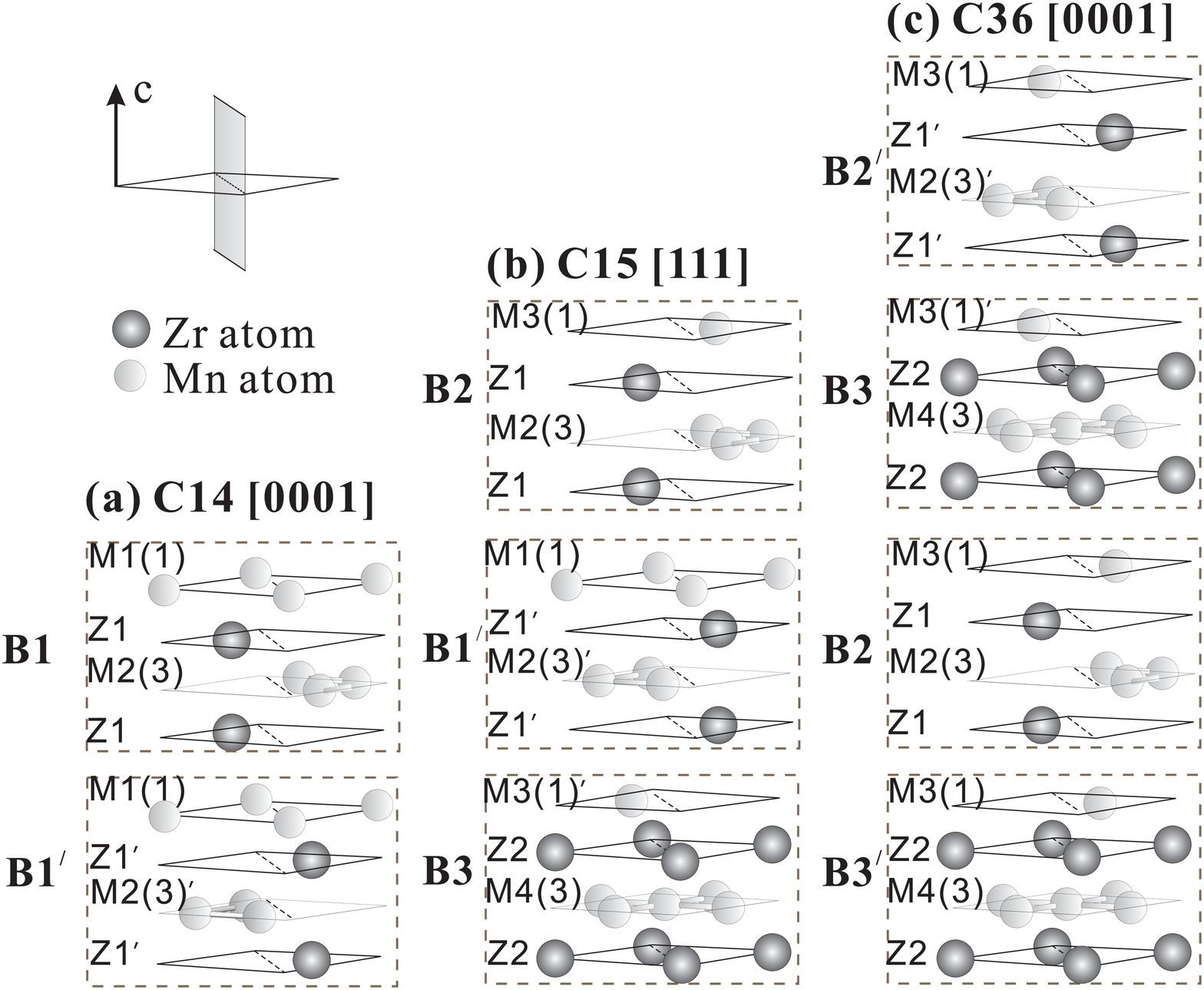}\\
\caption{Schematic stacking of Zr  and Mn layers for the hexagonal C14, C36
and cubic C15 crystal structures along the directions [0001] and [111],
respectively.  C14 (panel a) is build by the two blocks B1 and
B1$^{\prime}$.  B1 consists of  layers Z1 with one Zr atom per layer unit cell, and of
two planes M1(1) and M2(3).  M1(1) has 1 Mn atom per layer unit cell,
whereas the unit cell of layer M2(3) contains three Mn atoms
arranged in triangles typical for the structures of Laves phases.
Primed objects (blocks or planes) are obtained when mirroring the corresponding
unprimed object by the mirror plane shown in the insert (left top).  The C15
structure (panel b) is built by the three blocks B2, B1$^\prime$, and B3.  
Block B2 contains the Mn layer M3(1), and in block B3 the layer Z2 has Zr atoms
of weight 1/4 at the corners.  The layer M4(3) comprises  3 Mn atoms
per layer cell (Mn atoms on boundary lines have weight 1/2, the atom
in the center counts fully).  The C36 structure (panel c) is constructed by
stacking of blocks B2$^\prime$, B3, B2, and B3$^\prime$.
\label{fig1}}                                            
\end{center}
\end{figure}

Polymorphism of a solid material is its property to exist in several or even
many crystal modifications which are called polytypes.  A very famous example
of a polymorphic material is SiC \cite{starke1} for which the polytypes are
distinguished by the variations of stacking of SiC bylayers.  ZnS, as well,
appears in many polytypes \cite{zns1}. 
First principles calculations (as usual referring to T=0K)
on SiC \cite{heine1} and on ZnS and related II-VI compounds \cite{zns2}
revealed that the occurence of polytypism must be due to the energetical
degeneracy of structures at very low temperatures.  For Laves phase compounds
\cite{laves} so far only high temperature polymorphism is known from
experiments  on e.g.  TiCr$_2$, ZrCr$_2$, and HfCr$_2$ \cite{zrmn2p}.  
The polytypes are the most common structures of Laves phases compounds,
namely the densely packed hexagonal C14 and C36, and the cubic C15 crystal
structures which are distinguished by the stacking of layers and blocks as sketched in
Fig. \ref{fig1} for ZrMn$_2$.  Searching for the ground state structure of
ZrMn$_2$ we performed density functional calculations for the C14, C15 and C36
structures finding them energetically nearly degenerate at very low
temperatures.  This new finding, the first case of low temperature
polymorphism of a Laves phase compound, is the subject of our paper.
According to our results magnetism plays an important role in
the polymorphism of ZrMn$_2$.  Bulk samples of ZrMn$_2$ were fabricated so far
only at high temperatures of about 870 K at which the C14 structure was
observed \cite{nat2,zrmn2p}.  Our calculations confirm the stabilization of
this high temperature phase due to the free energy of the lattice
vibrations.  

We predict that at least the C14, C15 and C36 structures are
nearly equally stable at very low temperatures.  All three phases are
ferromagnetic and their structure dependent gain in magnetization energy is
at
just the right size to cause the degeneracy. The C15 structure, which is
the
most stable one at T=0K, is predicted to transform at T$_{tr}$ $\approx$ 160K to
the C14 structure as observed experimentally at elevated temperatures.  

Another type of structural degeneracy at low temperatures might occur due
to martensitic instabilities by which energetically closely related structures
are transformed into each other because of  elastic or vibrational
instabilities \cite{martensitic,parlinski1}.  
For ZrMn$_2$, however, our calculations reveal
no martensitic instability in the analysis of elastic
and vibrational properties which strengthens our argument that ZrMn$_2$ is a
truly polymorphic material.

Fig. \ref{fig1} schematically shows the construction of the C14, C36 and C15
crystal structures in terms of stacking of blocks consisting of Zr and Mn
layers.  As is visualized by the figure and described in the caption, the
stacking is such that blocks can only be combined when the layers are arranged
appropriately.  Following this recipe one might construct longer ranged
stacking with larger unit cells \cite{laves} which possibly might lead to a
large number of polytypes.

For the calculation of ground state properties of ZrMn$_2$ in the
C14, C15 and C36 crystal structures we applied density functional
theory (DFT) by means of the plane wave Vienna Ab initio
Simulation Package (VASP)\cite{vasp} which -in its projector augmented wave
formulation
- is one of the most precise methods for calculating the energetics and
 electronic structure of solid matter within periodic boundary conditions.
For the many-body exchange-correlation interaction the generalized gradient
approximation of Perdew and Wang \cite{gga91} in combination with the
approach of Ref. \cite{vosko} for spin polarization was chosen.  The
calculations are free from any empirical parameters. Lattice parameters as
well as the atomic coordinates are determined by minimizing total energies and
atomic forces.  Care was taken that for each structure a sufficient number of
{\bf k} points for the Brillouin zone integration was chosen. Ferromagnetic as
well as some selected antiferromagnetic spin arrangements were considered.
Details will be described elsewhere \cite{chen3}.  In order to take into
account the temperature dependent effects of lattice vibrations, the
phonon
dispersion and density of states were calculated by the so-called direct
approach \cite{phonon} utilizing data of VASP calculations.  Furthermore,
elastic constants for the C14 and C15 structure were determined from DFT energy
densities corresponding to suitable distortions.  The energy of formation was
obtained from the difference of total energies for the compound and the pure
phases \cite{chen1}.

Figure 2 contains the essential information of our study.  
Viewing  the nonmagnetic
(NM) results, clearly  the energy of the C14 structure is lowest for all volumes.  At
equilibrium volume $V_0$ corresponding to the minimum of the energy curves C14 is favoured by
about 1.5 kJ mol$^{-1}$ compared to C36, and by the substantially larger
amount of 5.3 kJ mol$^{-1}$
compared to  C15.  The important role of magnetism 
is already illustrated by the results for the most stable
cases of selected antiferromagnetic (AF) spin arrangements \cite{chen1,chen3}.
For all structures, the minimum energies are lowered, lying now much closer
together as compared to the NM case, with C36 as the favored structure. Also
noticeable is that the equilibrium volumes for C14 and C15 are quite different
expressing different magneto-volume effects.  The most striking result is
presented by the ferromagnetic (FM) data. Again, the energy is lowered for all
structures when compared to AF, but now C14, C15, and C36 nearly coincide. By the
very small energy difference of 0.3 kJ mol$^{-1}$ (corresponding to
$\approx$ 30K) the cubic C15 phase is now more stable than the two
hexagonal cases. At larger volumes, differences are getting slightly larger,
whereas at
smaller volumes (i.e. under pressure) the energy curves of all
three structures are becoming undistinguishable.

\begin{figure}
\begin{center}
 \includegraphics[width=85mm]{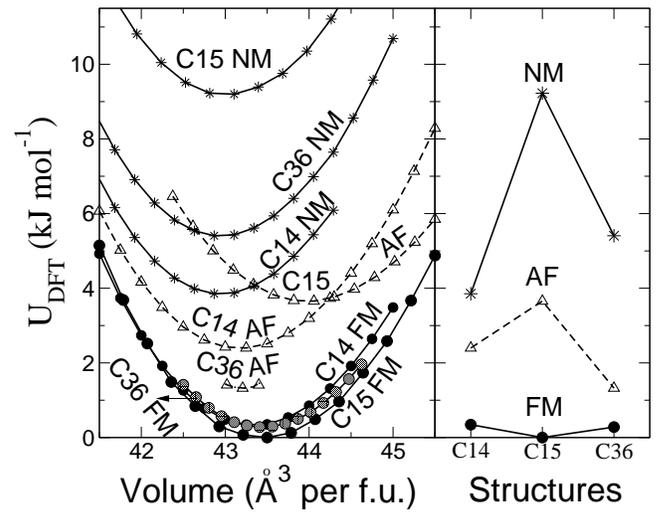}
 \caption{
Density functional results for ZrMn$_2$.  Left panel: Calculated energies
$U_{DFT}(V)$ as a function of volume $V$ for the nonmagnetic case (NM,
asterisks and full black lines), the most stable antiferromagnetic (AF,
triangles and dashed lines), and ferromagnetic (FM, dots and full  lines) spin
orderings for the C14, C15  and C36  structures of ZrMn$_2$.  Volume per
formula unit (3 atoms per unit cell).  For each marked point of the curves all
structural parameters are fully relaxed resulting in the minimum of total
energy for a given fixed volume.  Right panel: Energies $U_{DFT}$ at
equilibrium volume $V_0$ at minimum of the total energies of all the cases in
the left panel.  The zero of energy corresponds to lowest calculated energy
which was obtained for C15 FM.  \label{fig2}}
\end{center}
\end{figure}

The interpretation of these results is that for the crystallization process at
least at very low temperatures there is a choice between at least three
structures distinguishable by the stacking sequences according to Fig.
\ref{fig1}.  If one wants to fabricate low temperature polytypes of ZrMn$_2$,
molecular beam epitaxy seems to be the only choice.  Standard procedures of
bulk crystal growth are rather unsuitable for the creation of low temperature
polytypes, because at high temperatures ZrMn$_2$ crystallizes only in the C14
structure as observed at about 870K \cite{nat2}.  On cooling down the C14
structure might be frozen in as a metastable state. The thermodynamical
equilibrium, which according to our prediction is the C15 structure, might be
never reached in a finite time.  In addition, no simple deformation path can
be imagined to deform C14 to C15 or C36, making a structural transition even
less probable. At present, no unambigous experimental structural information
for low temperatures  is available.

For ZrMn$_2$, 
Stoner's theory predicts the occurence of band magnetism already
from the NM density of electronic states at Fermi energy.  The energetical
degeneracy of structures then occurs for the ferromagnetic (FM) case.  The
most important indication why this happens is revealed by the right panel of
Fig. \ref{fig2}.  The FM energies for the corresponding $V_0$ are almost lying
on one line quite in contrast to the NM case for which the C15 energy is rather
high compared to C14. By spontaneous FM spin polarization all three structures
gain different magnetization energies defined by $U_{DFT}(FM;V_0) -
U_{DFT}(NM;V_0)$, which are just of the right size to make the structures
degenerate.  The largest magnetization energy of 9.2 kJ mol$^{-1}$ is gained
in the cubic C15 structure.  The degeneracy clearly is due to an electronic
effect, since the structural changes, when relaxing geometries for all the
magnetic states,  have only insignificant influence on the energy differences.
The average magnetic moment of Mn in all cases is about 0.9 $\mu_B$.  It
should be noted, that for the hexagonal cases, not all Mn atoms are
equivalent, therefore also the magnetic moments differ significantly.  The
average magnetic moments, however,  agree very well between the
structures.  For the C14 structure the magnetic moments of the two non
equivalent Mn atoms are 1.24 and 0.60 $\mu_B$, for C15 the Mn moment is 0.90
$\mu_B$, and for the C36 case three non equivalent Mn atoms appear with
moments of 1.14, 0.92 and 0.71 $\mu_B$.  In contrast to other compounds with
long range atomic stacking sequences  \cite{cuauzral}, which are non magnetic,
for ZrMn$_2$ the ferromagnetic spin ordering for the energetical degeneracy of
suitable atomic stackings is crucial.

Concerning  antiferromagnetic spin orderings, it might be possible that more
complicated AF structures (including non collinear spin arrangements) might
lower the AF total energy.  However, also due to the rather small magnetic
moment we would not expect any substantial change of our findings that FM is
the most stable phase.  In any case, the structural degeneracy for FM ZrMn$_2$
would be unaffected.

\begin{figure}
\begin{center}
 \includegraphics[width=86mm]{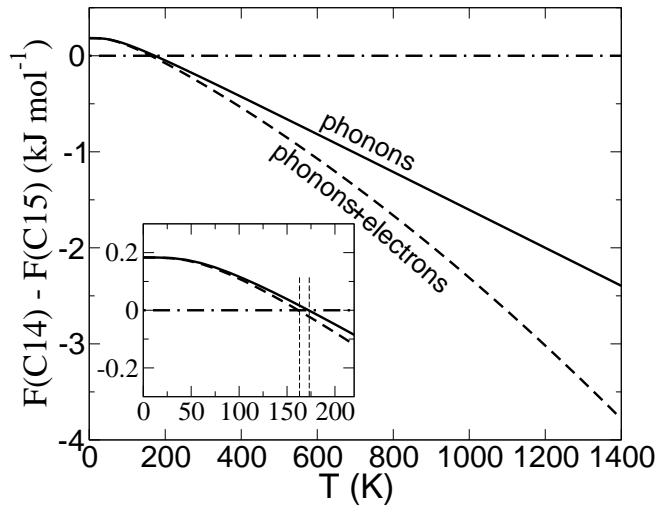}
 \caption{
Difference of free  energies $F(C14) - F(C15)$ of ZrMn$_2$
in the ferromagnetic state.
Shown are results with contributions of lattice vibrations only
(full line) and also with including electronic contributions (dashed line).
Positive values: C15 structure is preferred; negative values: C14 is stable.
Insert magnifies region around the temperature
T$_{tr}$ of the structural transition C15 $\rightarrow$ C14.
\label{fig3}}
\end{center}
\end{figure}

So far, the polymorphism of ZrMn$_2$ is a predicted property based on
calculated data.  The question is how reliable are the results derived by the
applied ab initio approach?  An established fact is the occurence of the C14
structure at elevated temperatures, which can be explained by our theoretical
approach on taking into account the lattice vibrations. From the calculated phonon
dispersion relations and phonon densities of states for C14 and C15 the temperature
dependent free vibrational energy $F_{vib}(T)$  is derived. Based on the
difference curves in  Fig.  \ref{fig3} we predict a structural phase
transition C15 $\rightarrow$ C14 at a transition temperature of T$_{tr}
\approx$ 160K.  A temperature dependent free energy contribution for the
electrons was included as well \cite{ashcroft}, which, however, is unimportant at T =
T$_{tr}$.  For temperatures larger than T$_{tr}$ C14 is stabilized at least in
comparison to the C15 structure. (Because of its large unit cell we did not
calculate the phonon dispersion relations for C36) Decomposing the free energy $F_{vib}
= U_{vib} - T  S_{vib}$ in contributions due to the internal energy $U_{vib}$
and the entropy  $S_{vib}$ we find that the C15 $\rightarrow$ C14 transition
is caused by the vibrational entropy.  Already the zero point
energy  $F_{vib}(T=0)$ contribution reduces the energy difference between C14 and C15.
Although the C14 and C15 structures are nearly energetically degenerate no
indication of any structural instability is seen in the phonon dispersion
relations \cite{chen3} as would be typical for martensitic
transformations \cite{martensitic}. 
Finally, the elastic constants
calculated for the C14 and C15 structures are typical for stable Laves phase compounds
\cite{chen3} and from the conditions for mechanical stability against any homogeneous elastic
deformations \cite{elastic_analysis}
such instabilities can be excluded for both structures.

The reliablity of our calculational approach is strongly confirmed by results
on related compounds. For ZrCr$_2$ \cite{zrcr2} and ZrFe$_2$ \cite{zrfe2}
recent experiments find the C15 structure stable up to temperatures close to
the melting point. When studying the three structures for these two compounds
we also find that the C15 structure is definitely preferred \cite{chen3}.
Interestingly, according to our calculations ZrCr$_2$ is nonmagnetic but
ZrFe$_2$ is found to be strongly ferromagnetic.  Analogous to ZrMn$_2$ the C15
structure gains significantly more magnetization energy as compared to the
hexagonal cases. Because for ZrFe$_2$ the magnetic moment of 1.9 $\mu_B$
(derived from our calculation as well as from experiment) is significantly
larger than for ZrMn$_2$ the magnetic energy gain of the C15 structure is also
enhanced and C15 is now stabilized. For ZrMn$_2$, as discussed above, the
magnetization energy leads to the degeneracy of structures. For TiMn$_2$ our
ab initio approach yields the correct C14 structure ground state \cite{chen1}.
The compound TiMn$_2$ is just nonmagnetic (C14) or weakly magnetic (C15).

An indicative quantity about the strength of bonding is the energy of
formation $F_{form} = F_{comp}(ZrMn_2) - F_0(Zr) - 2 F_0(Mn)$  defined by  the
differences of free energies at the equilibrium volumes for the compound,
$F_{comp}$, and the pure elements, $F_0$, respectively.  At zero temperature
the expression reduces to the differences of the corresponding energies
$U_{DFT}$ neglecting the weak temperature dependence for moderate temperatures
sufficiently lower than the melting temperature. From the differences between the
corresponding energies, $U_{DFT}$, we arrive at a formation energy of about -64
kJmol$^{-1}$ for the ferromagnetic ground state, which characterizes a rather
strong bonding between the Zr and Mn atoms. Therefore, from the formation
energy no indication of any phase instability can be inferred. The available
experimental information of -48 kJmol$^{-1}$ in Ref.  \cite{zrmn2eform} 
needs to be checked by more recent experiments.  The strong
formation energy of ZrMn$_2$ is quite in contrast to the very closely related
compound YMn$_2$.  From standard DFT calculations we derive even slightly
positive formation energies, and experimentally the compound hardly forms
\cite{chen3,ymn2}.  The reliablity of our calculated energies of formation is
confirmed by the results for TiMn$_2$ of about -90 kJmol$^{-1}$, on which
experiment and theory agree very well\cite{chen1}.

For the C14 structure of ZrMn$_2$, for which experimental data exists, the
calculated structural and lattice parameters are in good agreement with
experiment \cite{chen3}.  The calculated volume is slightly smaller by about
3.4\% possibly due to the approximations made for the many-body term of
DFT. It should, however, be noted that the deviations are small and that our
basic findings of energy degeneracy involve {\em differences} of energy, each of
them obtained with the same approximations.  Furthermore, the qualtity of our
data is corroborated by the calculated energies of formation which in many
cases agree with reliable experimental data.  Very recently, ab initio DFT
calculations for the C14 and C15 structures of ZrMn$_2$ were published
\cite{zrmn2fu}. We find perfect agreement for the magnetic moment but -due to
the lack of published data- no comparison of structural energy differences is possible.

Summarizing our results, based on density functional calculations we predict
ZrMn$_2$ to be polymorphic at low temperatures.  We propose the
degeneracy of at least three crystal structures (C14, C15, C36) which are
distinguishable by different stacking of layers. This degeneracy occurs in the
ferromagnetic phases, which are lowest in energy compared to nonmagnetic or
selected antiferromagnetic spin orderings. The gain in magnetization energy
due to spontaneous spin polarization is different for all investigated
structures but just of the right size to cause the degeneracy. The C15
structure is stable by a very small energy difference which might disappear
when pressure is applied. Including the free energies of lattice vibrations we
predict a structural phase transition at 160 K to the C14 structure which is
found experimentally at elevated temperatures.  The structural degeneracy is
not due to any instability: the phonon dispersions for the C14 and C15
structures show no soft modes, the energy of formation indicates strong
bonding, and the elastic constants for C14 and C15 exclude any weakness due to elastic
deformations. Finally, the reliability of our calculations is confirmed by
comparison to reliable experimental data, for related compounds as well.

This work was supported by the Austrian Science Fund FWF project nr. 14761. Most of
the calculations were performed on the Schr{\"o}dinger-2 PC cluster of the
University of Vienna.  X.Q.C. is grateful to the OEAD for support within the
Austrian Chinese technical scientific exchange program, project IV.A.15.

\end{document}